\documentclass[acmlarge]{acmart}
\renewcommand\footnotetextcopyrightpermission[1]{} 
\usepackage{enumerate}
\usepackage{hyperref}
\hypersetup{
  colorlinks=true,   
  linkcolor=red,          
  citecolor=blue,        
  filecolor=magenta,      
  urlcolor=cyan,          
  pagebackref=true, 
}
\usepackage{relsize}
\AtBeginDocument{%
  \providecommand\BibTeX{{%
    \normalfont B\kern-0.5em{\scshape i\kern-0.25em b}\kern-0.8em\TeX}}}

\setcopyright{acmcopyright}
\copyrightyear{2020}
\acmYear{2020}
\acmDOI{???}

\newcommand{\myTail}{
 \bibliographystyle{ACM-Reference-Format}
	 \bibliography{local} 
 \end{document}
}

\newcommand{\dt}[1]{{\bf #1}}
\newcommand{\ignore}[1]{}	

\newcommand{\mt}{\rightarrow}

\newcommand{\beq}{\begin{equation}}
\newcommand{\eeq}{\end{equation}}
\newcommand{\beql}[1]{\begin{equation}\label{eq:#1}}
\newcommand{\eeql}{\end{equation}}
\newcommand{\bemultl}{\begin{multline}}
\newcommand{\emultl}{\end{multline}}
\newcommand{\beqarrays}{\begin{eqnarray*}}
\newcommand{\eeqarrays}{\end{eqnarray*}}

\newcommand{\blem}{\begin{lemma}}
\newcommand{\elem}{\end{lemma}}
\newcommand{\bleml}[1]{\begin{lemma} \label{lem:#1}}
\newcommand{\eleml}{\end{lemma}}
\newcommand{\blemT}[2]{\begin{lemma}[#1] \label{lem:#2}}
\newcommand{\elemT}{\end{lemma}}
\newcommand{\bthm}{\begin{theorem}}
\newcommand{\ethm}{\end{theorem}}
\newcommand{\bthml}[1]{\begin{theorem} \label{thm:#1}}
\newcommand{\ethml}{\end{theorem}}
\newcommand{\bthmT}[2]{\begin{theorem}[#1] \label{thm:#2}}
\newcommand{\ethmT}{\end{theorem}}
\newcommand{\brem}{\begin{remark}}
\newcommand{\erem}{\end{remark}}
\newcommand{\breml}[1]{\begin{remark} \label{rem:#1}}
\newcommand{\ereml}{\end{remark}}
\newcommand{\bcor}{\begin{corollary}}
\newcommand{\ecor}{\end{corollary}}
\newcommand{\bcorl}[1]{\begin{corollary} \label{cor:#1}}
\newcommand{\ecorl}{\end{corollary}}
\newcommand{\bpropl}[1]{\begin{proposition} \label{pro:#1}}
\newcommand{\epropl}{\end{proposition}}

\newcommand{\bpf}{\begin{proof}}
\newcommand{\epf}{\end{proof}}

\newcommand{\refeq}[1]{(\protect\ref{eq:#1})}

\newcommand{\refLem}[1]{Lemma~\protect\ref{lem:#1}}
\newcommand{\refThm}[1]{Theorem~\protect\ref{thm:#1}}
\newcommand{\refFig}[1]{Figure~\protect\ref{fig:#1}}

\newcommand{\refSec}[1]{Section~\protect\ref{sec:#1}}

\newcommand{\refPro}[1]{Proposition~\protect\ref{pro:#1}}
\newcommand{\CC}{{\mathbb C}}
\newcommand{\NN}{{\mathbb N}}

\newcommand{\ZZ}{{\mathbb Z}}

\newcommand{\as}{\textcolor{red}{\mathop{\mbox{\rm :=}}}}
\newcommand{\dd}{ ,\ldots , }

\newcommand{\ib}{\subseteq }

\newcommand{\abs}[1]{\left\lvert#1\right\rvert}       

\newcommand{\calG}{\mathcal{G}}

\newcommand{\set}[1]{\left\{ #1 \right\}}

\newcommand{\paren}[1]{\left( { #1 }\right)}	
\newcommand{\su}{\cup}

\newcommand{\subsT}[1]{\biggr\rvert_{#1}}
\newcommand{\ceil}[1]{\left\lceil {#1} \right\rceil}

	\newcommand{\vfigpdf}[3]{
            \begin{figure}[htb]
              \centering
              \includegraphics[scale=#3]{#2}
              \caption{#1}
              \label{fig:#2}
            \end{figure}
          }
 \newcommand{\mmat}[2][ccccccccccccccccccccccccc]{\left[
		\begin{array}{#1}
		#2\\
		\end{array}\right]}

	\newcommand{\bfalpha}{{\boldsymbol{\alpha}}}

	\newcommand{\bfbeta}{{\boldsymbol{\beta}}}
	\newcommand{\bfmu}{{\boldsymbol{\mu}}}

	\newcommand{\bfq}{{\boldsymbol{q}}}

	\newcommand{\bfr}{{\boldsymbol{r}}}
	\newcommand{\calv}{{\mathcal{V}}}
	\newcommand{\whatf}{{\widehat{f}}}

\newcommand{\sdisc}{\text{sDisc}}
\newcommand{\mpdiff}[2]{ \frac{\partial^{#1} }{
                        \partial #2 ^{#1}}}	
\newcommand{\In}{\text{In}}
\newcommand{\wmax}{w_{\max}}
\renewcommand{\mod}{\text { mod }}
\newcommand{\sep}{\text{sep}}
\newcommand{\res}{\text{res}}
\begin{document}

\title{Generalizing The Davenport-Mahler-Mignotte Bound -- The Weighted Case} 
%
\author{Vikram Sharma}
\email{vikram@imsc.res.in}
\affiliation{%
  \institution{Institute of Mathematical Sciences, HBNI}
  \streetaddress{}
  \city{Chennai}
  \country{India}
  \postcode{600113}
}

\begin{abstract}
Root separation bounds play an important role as a complexity measure in understanding the behaviour
of various algorithms in computational algebra, e.g., root isolation algorithms. 
A classic result in the univariate setting is the Davenport-Mahler-Mignotte (DMM) bound.
One way to state the bound is to consider a directed acyclic graph $(V,E)$ on a subset of roots of a degree $d$
polynomial $f(z) \in \CC[z]$, where the edges point from a root of smaller absolute
value to one of larger absolute, and the in-degrees of all vertices is at most one. 
Then the DMM bound is an amortized lower bound on the following product:
$\prod_{(\alpha,\beta) \in E}|\alpha-\beta|$. However, the lower bound involves the discriminant of the polynomial
$f$, and becomes trivial if the polynomial is not square-free. This was resolved by Eigenwillig, (2008), 
by using a suitable subdiscriminant instead of  the discriminant. 
Escorcielo-Perrucci, 2016, further dropped the in-degree constraint on the graph by using
the theory of finite differences. Emiris et al., 2019, have generalized their result to
handle the case where the exponent of the term $|\alpha-\beta|$ in the product is at most the
multiplicity of either of the roots. In this paper, we generalize these results
by allowing arbitrary positive integer weights on the edges of the
graph, i.e., for a weight function $w: E \mt \ZZ_{>0}$, 
we derive an amortized lower bound on 
$\prod_{(\alpha,\beta) \in E}|\alpha-\beta|^{w(\alpha,\beta)}$. Such a product occurs in the 
complexity estimates of some recent algorithms for root clustering (e.g., Becker et al., 2016),
where the weights are usually some function of the multiplicity of the roots. 
Because of its amortized nature, our bound
is arguably better than the bounds obtained by manipulating existing results to accommodate
the weights. 
\end{abstract}



\maketitle

\section{Introduction}
\label{sec:intro}
Given a {\em monic} univariate polynomial $f(z) \in \CC[z]$, of degree $d$ with roots 
$\alpha_1 \dd \alpha_d$, not all distinct, a \dt{root separation bound} is a lower bound on
the smallest distance $\sep(f)$ between any distinct pair of roots of $f$. A 
classic result \cite{mignotte:bk} states that 
        $$\sep(f)  > d^{-(d+2)/2} \Delta(f)^{1/2} M(f)^{1-d},$$
where 
        $$\Delta(f) \as \prod_{i< j}(\alpha_i -\alpha_j)^2$$ 
is the \dt{discriminant} of $f$, and 
        \beql{mahler}
        M(f) \as \prod_{i=1}^d \max\set{1,|\alpha_i|}
        \eeql
is the \dt{Mahler measure} of $f$.

The parameter $\sep(f)$ naturally occurs in the complexity analysis of many algorithms; examples are
the (real or complex) root isolation algorithms (\cite{pan}, \cite{davenport}, \cite{esy:06}, \cite{becker+3}). 
However, most of these algorithms need a lower bound on 
the product of certain pairs of roots and not just the worst case separation. 
To capture these pairs, we consider a simple (i.e., no loops and
multiple edges) undirected graph $G=(V,E)$, whose vertices are a subset of the distinct roots of $f$.
Then we want a lower bound on $\prod_{(\alpha_i,\alpha_j) \in E}|\alpha_i - \alpha_j|$. 
One straightforward lower bound is $\sep(f)^{|E|}$,
but Davenport \cite{davenport} used the amortized nature of the Mahler measure
to derive a lower  bound for real roots that essentially matches the lower bound on $\sep(f)$ given above;
the argument was later modified by Mignotte to complex roots \cite{mignotte:95}. 
A consequence of these results is a straightforward improvement
in the complexity bounds on the running time of algorithms for root isolation algorithms 
by a multiplicative factor of the degree.

Both these lower bounds,  nevertheless, rely 
on the discriminant $\Delta(f)$ and are trivial when the polynomial is not square-free, i.e., 
it has multiple roots. A remedy is to work with the square-free part $\whatf$ of $f$,
but this again blows the bound by exponential factors because of the growth in the
coefficients of $\whatf$ as compared to $f$.
An alternative was presented by Eigenwillig \cite{eigenwillig} that uses the $(d-r)$-th sub-discriminant of $f$
instead of the the discriminant, where $r$ is the number of distinct roots of $f$. 
However, there are some constraints on the graph $G$ for the bound to be applicable, namely, 
in the directed acyclic graph obtained by directing the edges of $G$
from a root of smaller absolute value to one of larger absolute value, 
the in-degree of all the vertices is at most one.
Escorcielo-Perrucci \cite{perruci} dropped
this in-degree constraint by using the theory of finite differences. Despite this,
their result gives weaker bounds on products of the form 
        \beql{prodweight}
        \prod_{(\alpha_i,\alpha_j) \in E}|\alpha_i - \alpha_j|^{w(\alpha_i,\alpha_j)},
        \eeql
where $w: E \mt \NN$ is a \dt{weight function} that assigns a positive integer to all the edges.\footnote{
Throughout, we use $\NN$ to denote the set of positive integers and $\ZZ_{\ge 0}$ the set of non-negative
integers.}
In the special case where the weight function is such that the  
weight of an edge is bounded by  the multiplicity of one of its vertices,
\cite{ks} and \cite{emt} have derived lower bounds when the coefficients of $f$ are
real and complex numbers, respectively. To state their bound, let
$f$ have $r$ distinct roots $\alpha_1 \dd \alpha_r$ with multiplicities $m_1 \dd m_r$,
respectively, $\whatf$ denote the square-free part of $f$, and for a root
$\alpha_i$ let $\Delta_i$ denote the distance to the nearest distinct root. Then the bound
in \cite{emt} is the following: If $K \ib [r]$ and $w_i \in \NN$ is such that $w_i \le m_i$, for $i \in K$,
then
        \beql{emt}
          \prod_{i \in K} \Delta_i^{w_i} \ge 
                                2^{-d(r+2)}(\|f\|_\infty \|\whatf\|_\infty)^{-d} M(f)^{1-r}|\res(f,\whatf')|,
        \eeql
here $\|\cdot\|_\infty$ is the maximum absolute value over the coefficient sequence of the polynomial,
and $\res(\cdot,\cdot)$ is the univariate resultant. 
These bounds, though useful, fail to provide amortized lower bounds
when the $w_i$'s exceed the multiplicity. Such a scenario, for instance, 
occurs in the complexity analysis of some recent root clustering algorithms \cite{becker+3,bs}, where 
the following  product occurs, for some subsets $K_i \ib [r]$:
        $$\prod_{i \in K}\Delta_i^{\sum_{j \in K_i}m_j}.$$ 
One way to derive a lower bound on this product
is to exponentiate the left-hand side of \refeq{emt} to the degree $d$ (since the sum of the multiplicities
over $K_i$ is bounded by $d$), 
move the extraneous factors to the denominator in the right-hand
side, and upper bound these to get a lower bound on the desired product. But, just as was the case 
with $\sep(f)^{|E|}$ earlier, such an approach loses the amortization property 
and gives exponentially worse bounds.

In this paper, we derive a lower bound on the product in \refeq{prodweight} 
for arbitrary weight functions. The restrictions on the weights in the earlier approaches
was an outcome of the choice of the symmetric function (either the discriminant, sub-discriminant or the 
resultant). We instead choose a symmetric function based on the 
weights and try to optimize over all valid choices of the function. 
This is done by constructing a confluent Vandermonde matrix to 
get the desired weight structure in the exponents. 
The choice of the confluent Vandermonde is especially helpful when the
weights are skewed in distribution, because this means we can pick a different multiplicity structure
on the roots and obtain better bounds. The spectral structure of the weighted adjacency matrix 
$A_w \as [w_{i,j}]_{i,j=1\dd r}$ plays an important role in the choice of the multiplicity structure
for constructing the confluent Vandermonde matrix. For ease of comprehension,
we state our result when $f$ is an  integer polynomial (since then the absolute value of
the non-zero symmetric function is at least one, which is how the bounds are used in practice)
and is also monic (otherwise divide $M(f)$ by the absolute value of the leading coefficient).
Let $\|A_w\|_\star$ denote the \dt{nuclear norm} of $A_w$, i.e., the sum of its singular values,
$n \as r\ceil{\sqrt{\|A_w\|_\star}}$, and
$w(E)$ be the sum of the weights over the edges of $G$. Then we show that
\beql{result1}
  \prod_{(\alpha_i,\alpha_j) \in E}|\alpha_i - \alpha_j|^{w(\alpha_i,\alpha_j)} >
        M(f)^{-2r\|A_w\|_\star} \paren{\frac{n}{\sqrt{3}}}^{-\frac{3r\|A_w\|_\star}{2}-w(E)}n^{-n/2}.
 \eeql
The bound is amortized because the exponent of the Mahler measure does not contain $w(E)$,
which would be the case if we try to derive the lower bound by modifying the earlier results
(see \refeq{perrucic} below).

In the next section, we give the requisite details and properties of the confluent Vandermonde
matrix; \refSec{main-result} contains the statement of our main result \refThm{main} and its comparison
with a modification of an existing bound; \refSec{proof} contains a proof of the main result,
and in \refSec{best-matrix} we  specialize it to obtain the form given above in \refeq{result1}. 
       
\section{Confluent Vandermonde}
\label{sec:confl-vand}
Consider the column vector
        $$v(x)^t \as \mmat{1 & x & x^2 & \cdots & x^n}.$$
Define the vector obtained by differentiating each entry in the column above $i$ times and dividing by $i!$, ie.,
        \beql{vi}
        v_i(x)^t  \as \mmat{{0\choose i} x^{-i} & {1\choose i} x^{1-i} &  {2\choose i} x^{2-i} & \cdots &{n-1\choose i} x^{n-1-i}},
        \eeql
with the natural convention that ${j \choose i} = 0$ if $j < i$.
Let 
        $$\bfbeta \as(\beta_1 \dd \beta_r) \in \CC^r$$ 
be an $r$-dimensional vector of complex numbers,
        $$\bfmu \as (\mu_1 \dd \mu_r) \in \NN^r$$ 
be a sequence of positive integers, and $n \as \sum_i \mu_i$. 
Then the \dt{confluent Vandermonde matrix} 
$V(\bfbeta;\bfmu)$ is the $n \times n$ matrix with columns $(v_j(\beta_i))$, where
$1\le i \le r$ and $0 \le j \le \mu_i-1$. We will also use the notation $V(\beta_1 \dd \beta_r; \mu_1 \dd \mu_r)$
when we want to emphasize the $\beta_i$'s and $\mu_i$'s.  
We illustrate it below for $r=3$ and $\mu_1=2,\mu_2=3$.
        $$V(\bfbeta,\bfr) = \mmat{1 & 0                                               & 1 & 0 & 0\\
                                \beta_1&1                                       &\beta_2 & 1 &0\\                                  
                                \beta_1^2&{2 \choose 1}\beta_1  &\beta_2^2& 2\beta_2&1\\
                                \beta_1^3&{3 \choose 1}\beta_1^2  &\beta_2^3&3\beta_2^2&{3 \choose 2}\beta_2\\
                                \beta_1^{4}&{4 \choose 1}\beta_1^{3} &\beta_2^4&4\beta_2^3&{4\choose 2} \beta_2^2
                                }.$$
The \dt{block, $B(\beta_i)$, corresponding to a } $\beta_i$,
is the set of columns $(v_j(\beta_i))$, for $j=0 \dd \mu_i - 1$.
If all the $\mu_i$'s are one then we obtain the standard Vandermonde matrix denoted as 
$V(\bfbeta)$.
A key observation in understanding the determinant of the matrix above is to consider the matrix
obtained by replacing the last column $v_{\mu_i-1}(\beta_i)$, corresponding to some $\beta_i$ with $\mu_i > 1$, 
with the column $v(y)$, for some variable $y$, which gives us the matrix
        \beql{vy}
        V(y) \as  V(\beta_1 \dd \beta_i,y, \beta_{i+1} \dd \beta_r; \mu_1 \dd \mu_i-1, 1, \mu_{i+1} \dd \mu_r).
        \eeql
Let $\calv(y) \as \det(V(y))$.
By expanding along the column corresponding to $y$, we can express $\calv(y)$ as  a polynomial in $y$
with degree at most $n-1$. If we differentiate this polynomial $(\mu_i-1)$ times, divide by $(\mu_i-1)!$ and
substitute $y = \beta_i$, then we will recover the determinant of  $V(\bfbeta; \bfmu)$ expanded along
the last column of the block $B(\beta_i)$. More precisely, 
        \beql{vydiff}
        \det(V(\bfbeta; \bfmu)) = \frac{\calv^{(\mu_i-1)}(y)}{(\mu_i-1)!}\subsT{y=\beta_i}.
        \eeql
This result is crucial in deriving the following explicit form for the determinant \cite{hj}.
\bpropl{det}
The determinant of the confluent Vandermonde matrix satisfies
        $$\det(V(\bfbeta;\bfmu)) = \prod_{1 \le i < j \le r} (\beta_j - \beta_i)^{\mu_i\mu_j}.$$
\epropl

\section{The Davenport-Mahler-Mignotte Bound}
\label{sec:main-result}
The following variant of the bound appears in \cite{perruci}. 

\bpropl{perruci}
Let $\bfalpha \as (\alpha_1 \dd \alpha_r)$ be a sequence of distinct complex numbers, 
and 
        \beql{Malpha}
        M(\bfalpha) \as \prod_{i=1}^r \max\set{1, |\alpha_i|}.
        \eeql
If $G(V,E)$ is an undirected simple graph (i.e., with no multi edges and self-loops) 
with vertices $V \ib \set{\alpha_1 \dd \alpha_r}$,  then
        $$\prod_{(\alpha_i,\alpha_j) \in E}|\alpha_i -\alpha_j| \ge |\det(V(\bfalpha))| M(\bfalpha)^{-(r-1)}
                                        \paren{\frac{r}{\sqrt{3}}}^{-{|E|}}r^{-r/2}.$$
\epropl

{\bf Remark:} The result in \cite{perruci} actually uses the sub-discriminant.
Given a degree $d$ polynomial 
        $$f(z) = \prod_{i=1}^r(z-\alpha_i)^{m_i},$$
with distinct roots $\alpha_i$ of multiplicity $m_i$, for $1 \le i \le r$, the $(d-r)$ discriminant of $f$
is given by
        \beql{sdisc}
        \sdisc_{d-r}(f) \as\det(V(\bfalpha)) \prod_{j=1}^r m_i.
        \eeql
Taking absolute values and substituting the expression for the absolute value of the
determinant into \refPro{perruci} we get
        \beql{perruci}
        \prod_{(\alpha_i,\alpha_j) \in E}|\alpha_i -\alpha_j| \ge \sdisc_{d-r}(f)^{1/2}M(f)^{-(r-1)}
                                         \times\paren{\frac{r}{\sqrt{3}}}^{-{|E|}} \frac{r^{-r/2}}{\prod_{i=1}^r \sqrt{m_i}}.
         \eeql
Escorcielo-Perrucci \cite{perruci} then use the following upper bound
by Eigenwillig \cite{eigenwillig} to derive the final form of their result:
If $m_1 \dd m_r \in \NN$ and $\sum_{i=1}^r m_i = d$ then
        $$\prod_{i=1}^r \sqrt{m_i} \le 3^{\min\set{d, 2(d-r)}/6}.$$
Instead, if we use the AM-GM inequality then we get a sharper bound, namely
        $$\prod_{i=1}^r \sqrt{m_i} \le \paren{\frac{d}{r}}^{r/2}.$$
Substituting this in \refeq{perruci}, we get the following improvement over \cite{perruci}:
        $$
        \prod_{(\alpha_i,\alpha_j) \in E}|\alpha_i -\alpha_j| \ge \sdisc_{d-r}(f)^{1/2}M(f)^{-(r-1)}
                                        \paren{\frac{r}{\sqrt{3}}}^{-{|E|}}d^{-r/2}.
        $$

We will generalize \refPro{perruci}
above to account for non-zero integer weights on the edges, i.e., 
a lower bound on the product given in \refeq{prodweight}.
To illustrate the advantage of our approach,
we first give the details of a lower bound obtained by a straightforward modification
of \refPro{perruci}. 

Let $\wmax$ be the largest weight over all the edges in $G$. Then we can raise
the bound in the \refPro{perruci} to this weight and move the extraneous factor to the right-hand side,
and replace them with an upper bound. For any edge $(\alpha_i, \alpha_j) \in E$, we have
        $$|\alpha_i -\alpha_j|^{\wmax - w (\alpha_i, \alpha_j)} \le (2 M(f))^{\wmax}.$$
Therefore, we obtain the following lower bound as a modification of \refPro{perruci}, which we will
use to compare with the bound derived in this paper:
        \beql{perrucic}
          \prod_{(\alpha_i,\alpha_j) \in E}|\alpha_i -\alpha_j|^{w(\alpha_i,\alpha_j)} 
          \ge |\det(V(\bfalpha))|^{\wmax} M(\bfalpha)^{-((r-1)\wmax + |E|\wmax)} \cdot
           2^{-(|E|\wmax)} \paren{\frac{r}{\sqrt{3}}}^{-{|E|\wmax}}r^{-(r\wmax)/2}.
        \eeql
In comparison, we obtain the following generalization:
\bthml{main}
Let $\alpha_1 \dd \alpha_r \in \CC$ be distinct complex numbers.
Let $G(V,E)$ be an undirected graph whose vertices $V$ is a subset of $\set{\alpha_1 \dd \alpha_r}$,
with an associated a weight function $w: E \mt \NN$. Denote by 
        $$A_w=[w(\alpha_i,\alpha_j)]_{i,j=1\dd r}$$
the associated weighted adjacency matrix. To every vertex $\alpha_i \in V$,
we assign a potential $\mu_i \in \NN$ such that 
for every edge $(\alpha_i,\alpha_j) \in E$, we have $w(\alpha_i, \alpha_j) \le \mu_i \mu_j$.
Define $\bfmu$ as the column-vector of these potentials,
$n \as \sum_{i=1}^r \mu_i$, $M(\bfalpha)$ be as in \refeq{Malpha},
and $w(E)$ as the sum of the weights of the edges in the graph $G$, i.e., 
        \beql{we}
        w(E) \as \sum_{(\alpha_i,\alpha_j) \in E} w(\alpha_i,\alpha_j).
        \eeql
Then
        \beql{main}
          \prod_{(\alpha_i,\alpha_j) \in E}|\alpha_i -\alpha_j|^{w(\alpha_i,\alpha_j)}
          > |\det(V(\bfalpha;\bfmu)|\; M(\bfalpha)^{-\|\bfmu\bfmu^t - A_w\|_\infty}\\
           \quad\paren{\frac{n}{\sqrt{3}}}^{-\sum_i {\mu_i \choose 2}-w(E)}n^{-n/2},          
        \eeql
where $\infty$-norm of a matrix is the maximum one-norm over all the rows of the matrix. 
\ethml

{\bf Remarks:} 
\begin{enumerate}
\item Since we are dealing with symmetric matrices, we can replace the $\infty$-norm
  with the induced $1$-norm, which is the sum of the columns.
  
\item If all the weights are one, then we can take $\mu_i$'s as $1$, and obtain \refPro{perruci} as a corollary.

\item There is an interesting trade-off between the absolute values of
  the exponent of $M(\bfalpha)$ and $n/\sqrt{3}$, namely,
  as the number of edges  in $G$ increases the former decreases whereas the latter increases.
\end{enumerate}

In order to compare \refeq{perruci} and \refeq{main}, we 
make three assumptions:
\begin{enumerate}[(i)]
\item $G$ is connected, so $|E| \ge r-1$,
\item $\mu_i = \sqrt{\wmax}$, for all $i=1 \dd r$, and
\item $f$ is an integer polynomial. 

\end{enumerate}
From the last assumption, it follows that both $|\det(V(\bfalpha))|$ 
and $|\det(V(\bfalpha; \bfmu))|$ are at least one, and that is how we often use them in applications.
The second assumption implies that $n = r \sqrt{\wmax}$. We now compare three 
analogous terms from both the bounds by taking logarithms.

From the assumption of connectivity, it follows that the absolute
value of the exponent of $M(\bfalpha)$ in \refeq{perruci} is 
at least $2(r-1) \wmax$, whereas  in \refeq{main} it is at most $r\wmax$. If $r \ge 2$, 
then it follows that the former is larger than the latter. The
difference is because of the amortized property of the bound in \refeq{main}.

Consider the negation of the logarithm of the term 
        $n^{-\sum_i {\mu_i \choose 2}-w(E)}$
in \refeq{main}. This is equal to
        $$\paren{\sum_i {\mu_i \choose 2}+w(E)} \log n \le \paren{\sum_i {\mu_i \choose 2}+ |E|\wmax}\log (r\sqrt{\wmax}).$$
Since ${\mu_i \choose 2} \le \mu_i^2/2$, it follows that
$\sum_i {\mu_i \choose 2} \le r \wmax/2$. Therefore, the right-hand side above is upper bounded
by
        $$2|E|\wmax \log(r \sqrt{\wmax})$$
which is somewhat larger than $(-\log r^{-|E|\wmax})$, the corresponding term
in \refeq{perruci}. It must be remarked, nevertheless,  that the choice
in the second assumption is not the best (see \refSec{best-matrix}) and is only used for illustration at this point.

The negation of the logarithm of $n^{-n/2}$ in \refeq{main}
is 
        $$r\sqrt{\wmax} \log(r\sqrt{\wmax}),$$ 
which is better than the corresponding term in \refeq{perruci}, namely,
        $$(r\wmax) \log r,$$ 
for sufficiently large $\wmax$.

\subsection{Some Results from the Theory of Finite Differences}
Let $f: \CC \mt \CC$ be a function and $y_1 \dd y_n$ be $n$ nodes. Then the divided difference
of $f$ on these $n$ nodes is given by
        \beql{dd}
        f[y_1 \dd y_n] \as \sum_{k=1}^n \prod_{\ell=1; \ell \neq k}^n \frac{1}{(y_k- y_\ell)} f(y_k).
        \eeql
If $f(z) \as z^m$, for some $m \in \ZZ_{\ge 0}$, then we have the following closed form:
        \beql{closedf}
        f[y_1 \dd y_n] =
        \begin{cases}
          \mathlarger{\sum}_{\substack{(t_1 \dd t_n)\in \ZZ_{\ge 0}^n\\ 
                                        \sum_{i=1}^n t_i = m-n+1}} \prod_{j=1}^n y_j^{t_j} & \text{ if } n \le m+1\\
          0 & \text{ if } n > m+1.
        \end{cases}
        \eeql
Given $i_1 \dd i_n \in \ZZ_{\ge 0}$, denote by 
        \beql{pderiv}
        f^{(i_1 \dd i_n)}[y_1 \dd y_n] \as \frac{1}{i_1!}\mpdiff{i_1}{y_1} \cdots \frac{1}{i_n!}\mpdiff{i_n}{y_n}f[y_1 \dd y_n].
        \eeql
Then the following claim is straightforward to show:
\bleml{coeff}
Given $i_1 \dd i_n \in \ZZ_{\ge 0}$, the quantity 
        $$f^{(i_1 \dd i_n)}[y_1 \dd y_n]$$ 
is a linear combination
of $f^{(k_j)}(y_j)$, where $j = 1 \dd n$ and $k_j = 0 \dd i_j$. Moreover, the coefficient of
$f^{(i_j)}(y_j)$ in this linear combination is 
        $$\frac{1}{i_j!}\prod_{\ell=1: \ell \neq j}^{n} \frac{1}{(y_j-y_\ell)^{i_\ell+1}}.$$

\eleml
\bpf
For simplicity, we only argue for $i_1$; the argument is similar for other cases.
Consider the effect of $\frac{1}{i_1!}\mpdiff{i_1}{y_1}$ on $f[y_1 \dd y_n]$.
By linearity of the derivative operator, we only need to focus on the term 
$f(y_1)/\prod_{i\neq 1}(y_1-y_i)$.
From Leibniz's rule applied to this term we get the expression
        $$\frac{1}{i_1!}\frac{f^{(i_1)}(y_1)}{\prod_{i\neq 1}(y_1-y_i)}.$$ 
The effect of the other partial derivatives $\frac{1}{i_\ell!}\mpdiff{i_\ell}{y_\ell}$ is
only on the terms in the denominator, which yields the desired expression for the
coefficient of $f^{(i_1)}(y_1)$.
\epf

If $f(z) \as z^m$, for some $m \in \ZZ_{\ge 0}$, and $(i_1 \dd i_n) \in \ZZ_{\ge 0}^n$, 
then as a generalization of \refeq{closedf}, we obtain the following
        \beql{closedpf}
        f^{(i_1 \dd i_n)}[y_1 \dd y_n] =
        \begin{cases}
          \mathlarger{\sum}_{\substack{(t_1 \dd t_n)\in \ZZ_{\ge 0}^n\\ 
                                        \sum_{i=1}^n t_i = m-n+1}} \prod_{j=1}^n \mathsmaller{{t_j \choose i_j}y_j^{t_j-i_j}}
                                                                        &\mathsmaller{\text{\small{if }}n \le m+1}\\
                                    0 &\mathsmaller{\text{\small{if }}n > m+1}\\
        \end{cases}
        \eeql
with the natural convention that ${t_j \choose i_j} = 0$ if $t_j < i_j$.

\section{A proof of the main result} \label{sec:proof}
The idea of the proof is similar to \cite{perruci}. Given the undirected graph $G$, we first direct
its edges to go from a root of smaller modulus to one of larger modulus; this way we obtain a
directed acyclic graph $\calG$; the in-degrees of the vertices in $\calG$ can be larger than one, 
which is the case addressed in \cite{perruci}. We consider the vertices of $\calG$ in the reverse order of a topological sort on its vertices, i.e.,
in the order $(\alpha_1 \dd \alpha_r)$, where if $(\alpha_i,\alpha_j)$ is an edge in $\calG$ then $j < i$.
Let $\In(\alpha_i)$ denote the set of all vertices that have an edge pointing to $\alpha_i$, 
$d_i$ be the cardinality of $\In(\alpha_i)$ (i.e., the in-degree of $\alpha_i$),
and 
        \beql{V0}
        V_0 \as V(\bfalpha;\bfmu). 
        \eeql
At the $i$th step we will  process 
the block corresponding to $\alpha_i$ in $V_{i-1}$, where $i \ge 1$, to obtain a matrix $V_i$.
The relation between the two matrices is the following:
        \beql{vi1}
        \det V_{i-1} = \det (V_i) \prod_{\alpha_j \in \In(\alpha_i)} (\alpha_i - \alpha_j)^{w(\alpha_j, \alpha_i)}.
        \eeql   
The matrix $V_i$ is instead obtained from $V_{i-1}$
in stages by modifying the columns in the block corresponding to $\alpha_i$, that is,
there are two loops -- one over the blocks $B(\alpha_i)$, and an inner loop processing
the columns of the block $B(\alpha_i)$. The end result is a matrix $V_r$ such that
        $$\det(V_0 ) = \det(V_r) \prod_{i=1}^r \prod_{\alpha_j \in \In(\alpha_i)}(\alpha_i-\alpha_j)^{w(\alpha_j,\alpha_i)}.$$
The final step is to derive an upper bound on $|\det(V_r)|$; this is done by applying Hadamard's
inequality, and obtaining upper bounds on the two-norms of the columns of $V_r$.
In what follows, we will use
$\alpha$ in place of $\alpha_i$, $\mu_\alpha$ as the size of the block $B(\alpha)$, $k \as d_i$,
and $V \as V_{i-1}$.

Without loss of generality, let us assume that $\beta_{1} \dd \beta_{k}$ are the
$k$ vertices in $\In(\alpha)$, with respective weights $w_1 \dd w_k$.
Since we are processing the vertices in reverse topological order, we know that the blocks
corresponding to these vertices have not been changed. Let $\mu_1 \dd \mu_k$ be the sizes of the
blocks $B(\beta_1) \dd B(\beta_k)$, respectively.
We will replace each column in the block $B(\alpha)$
by a suitable linear combination of the columns in the blocks  
$B(\alpha)$ and $B(\beta_i)$ for $i=1 \dd k$. 
The linear combination will be obtained by taking a suitable partial derivative of the
form given in \refeq{pderiv} and then substituting $y_i$'s appropriately. Ideally, we would have
replaced, say the last column in $B(\alpha)$, by the partial derivative obtained by taking full weights,
$w_1 \dd w_k$. However, there is a slight obstacle, namely, that the derivatives of $f(\beta_i)$, for $i=1 \dd k$,
cannot exceed beyond $\mu_i-1$. To overcome this we assign each  
edge $(\beta_i, \alpha)$ with corresponding weight $w_i$ to a column in the block $B(\alpha)$,
namely to the $\ceil{w_i/\mu_i}$-th column in $B(\alpha)$; since $w_i \le \mu_i \mu_\alpha$ by assumption
on weights, the edge will be assigned to a column  in $B(\alpha)$. 
Let $S_j\ib [k]$, for $j=1 \dd \mu_\alpha$,  denote the set of all
indices assigned to the $j$th column of $B(\alpha)$, i.e.,
        \beql{sj}
        S_j \as \set{i \in [k]: \ceil{w_i/\mu_i} = j}.
        \eeql
By assignment it follows that $S_j$'s form a partition
of $[k]$. The reason why this assignment works is the following:
each column in $B(\alpha)$, along with its preceding columns in $B(\alpha)$
and the blocks $B(\beta_1) \dd B(\beta_k)$, can be used to factor out  $(\beta_i - \alpha)^{\mu_i}$;
therefore, $\ceil{w_i/\mu_i}$ columns will be required to get to $(\beta_i - \alpha)^{w_i}$. 
An illustrative aid for the subsequent proof is provided in \refFig{sketch}.

\vfigpdf{The matrix $V_{i-1}$ and the block $B(\alpha)$ at stage $i$ of the proof.
At the $j$th step in processing $V$, the columns $(j+1)$ to $\mu_{\alpha}$ of the block 
$B(\alpha)$ have been processed to obtain $V^{(j+1)}$. In $V^{(j+1)}$, the $j$th column is processed to 
obtain $V^{(j)}$.}{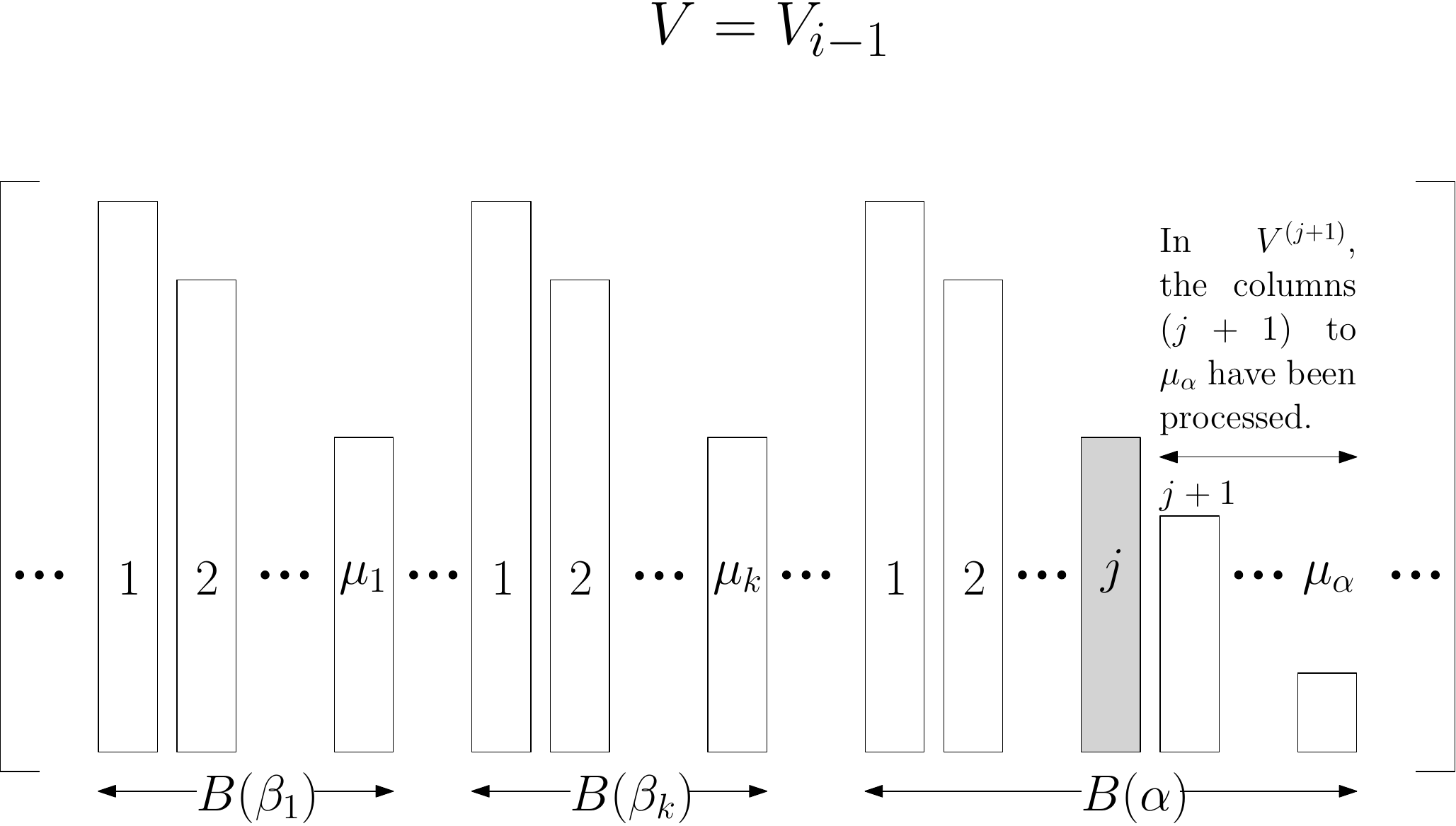}{0.4}

We will now process the columns of $B(\alpha)$ starting from {\em the last column to the first in $V$};
it will help the reader to note that the columns will be counted from $1$ to $\mu_\alpha$. 
Suppose we have already processed the columns of $B(\alpha)$ from $\mu_\alpha$
down to $(j+1)$  in $V$; let $V^{(j+1)}$ be the resulting matrix; 
initially, define $V^{(\mu_\alpha+1)} \as V$. 
For $\beta_\ell$, $\ell =1 \dd k$, define
        \beql{rl}
        r_\ell \as 
        \begin{cases}
          \mu_\ell & \text{ if $w_{\ell}$ is divisible by $\mu_\ell$,}\\
          (w_{\ell} \mod \mu_\ell) & \text{ otherwise}.
        \end{cases}
        \eeql
We inductively claim the following relation for $j \le \mu_\alpha$:
        \beql{hypo}
        \det(V) = \det(V^{(j+1)}) \prod_{\kappa =j+1}^{\mu_\alpha} \prod_{\ell \in S_\kappa} (\beta_\ell-\alpha)^{(\kappa-j-1)\mu_\ell+r_\ell}.
        \eeql
The proof is by reverse induction on decreasing values of $j$; the base case
trivially holds for $j= \mu_\alpha$, since the product vanishes and $V= V^{(\mu_\alpha+1)}$
by choice.

To complete the inductive claim \refeq{hypo}, we have to obtain the following terms
from the $j$th column in $B(\alpha)$:
\begin{enumerate}
\item the residue terms $(\beta_\ell - \alpha)^{r_\ell}$, for each index $\ell\in S_j$, and 
\item a factor of $(\beta_\ell - \alpha)^{\mu_\ell}$ for all the indices $\ell \in S_\kappa$, where $\kappa > j$.
\end{enumerate}
This is done by taking a suitable partial derivative of the finite difference. 
Let 
        \beql{Nj}
        N_j \as |\su_{\kappa = j}^{\mu_\alpha}S_\kappa|,
        \eeql
that is the total number of indices
assigned to column $j$ or greater; clearly $N_j \le k$. We will introduce $N_j$ variables
for each of these indices, and a variable $y_0$ for $\alpha$.
Note that the $j$th column of the block $B(\alpha)$ in $V^{(j+1)}$
is obtained by substituting $z = \alpha$ in $v_{j-1}(z)$ given in \refeq{vi}. The $m$th entry of this
column, for $m=1 \dd n$, is 
        \beql{mthentry}
        {m-1 \choose j-1} z^{m-j}= \frac{(z^{m-1})^{(j-1)}}{(j-1)!}.
        \eeql
Define $f_m(z) \as z^{m-1}$, and consider the finite difference  
        $$f_m[y_0, y_1 \dd y_{N_j}],$$ 
where the $y_\ell$'s are variables. 
Since the order of $y_i$'s in \refeq{pderiv} does not matter, we can assume without loss
of generality that  $S_j = \set{1 \dd |S_j|}$, the indices in $S_{j+1}$ are the next $|S_{j+1}|$
numbers, and so on $S_{\mu_{\alpha}}$ is the last $|S_{\mu_{\alpha}}$ numbers smaller than $N_j$;
thus the sets  $S_\kappa$, for $\kappa= j \dd \mu_\alpha$, form a partition of the
set $\set{1 \dd N_j}$. Further define
        \beql{i0}
        i_0 \as j-1,\; i_\ell \as r_\ell -1, 
        \eeql
for $\ell =1 \dd |S_j|$, and 
        \beql{il}
        i_\ell = \mu_\ell-1,
        \eeql
for $\ell = |S_j|+1 \dd N_j$. 
Then we replace the $m$th entry of the $j$th column $v_{j-1}(\alpha)$ in the matrix $V^{(j+1)}$ by
        \beql{fmi}
        f_m^{(i_0 \dd i_{N_j})}[y_0, y_1 \dd y_{N_j}]
        \eeql
and substitute $y_0 \as \alpha$, and $y_\ell \as \beta_\ell$, for $\ell=1 \dd N_j$.
This is done for all the $n$ entries (that is, $m=1 \dd n$) in the $j$th column. 
Let $V^{(j)}$ be the resulting matrix.
From \refLem{coeff} we know that the coefficient of $f_m^{(i_0)}(y_0)$ is 
        \beql{scaling}
        \frac{1}{i_0!}\prod_{\ell=1}^{N_j} \frac{1}{(y_0-y_\ell)^{i_\ell+1}}
                = \frac{1}{(j-1)!} \prod_{\kappa =j}^{\mu_\alpha} \prod_{\ell \in S_\kappa} \frac{1}{(\alpha-\beta_\ell)^{i_\ell+1}},
         \eeql
which is same for all $m=1 \dd n$. Therefore, the replacement 
of the entries of the $j$th column in the matrix $V^{(j+1)}$ by \refeq{fmi}, for $m=1 \dd n$,
is tantamount to obtaining the matrix $V^{(j)}$  from $V^{(j+1)}$ by replacing the $j$th
column of $V^{(j+1)}$ by a linear combination of its other columns and a scaled version of the $j$th column,
where the scaling factor is the product term in \refeq{scaling}; the $1/(j-1)!$ is not part of the scaling
as it already occurs in all the entries of the column (see \refeq{mthentry}). 
In terms of the determinant, we obtain  the following relation:
\begin{align*}
        \det(V^{(j+1)}) 
                &= \det(V^{(j)})
                          \prod_{\kappa =j}^{\mu_\alpha} \prod_{\ell \in S_\kappa} (\beta_\ell-\alpha)^{i_\ell+1}\\
                &= \det(V^{(j)}) \prod_{\ell \in S_j} (\beta_\ell-\alpha)^{r_\ell}
                        \prod_{\kappa =j+1}^{\mu_\alpha} \prod_{\ell \in S_\kappa} (\beta_\ell-\alpha)^{\mu_\ell}.
\end{align*}
Substituting this in \refeq{hypo}, we have the desired inductive relation:
        $$\det(V) = \det(V^{(j)}) \prod_{\kappa =j}^{\mu_\alpha} \prod_{\ell \in S_\kappa} (\beta_\ell-\alpha)^{(\kappa-j)\mu_\ell+r_\ell}.$$
We stop when $j=1$, to get
        $$\det(V) = \det(V^{(1)}) \prod_{\kappa =1}^{\mu_\alpha} \prod_{\ell \in S_\kappa} (\beta_\ell-\alpha)^{(\kappa-1)\mu_\ell+r_\ell}.$$
But recall from \refeq{sj} that for an index $\ell \in S_\kappa$, we have $\ceil{w_\ell/\mu_\ell} = \kappa$.
Furthermore, from \refeq{rl} it follows that
$w_\ell = (\kappa-1)\mu_\ell + r_\ell$. Since $\su_{\kappa=1}^{\mu_\alpha}S_\kappa= [k]$, we have accounted
for all the $\beta_\ell$'s, and so the equation above is the same as
        \begin{align*}
        \det(V) &= \det(V^{(1)}) \prod_{\ell=1}^k (\beta_\ell-\alpha)^{w_\ell}\\
                     &=\det(V^{(1)}) \prod_{\alpha_j \in \In(\alpha_i)} (\alpha_j-\alpha)^{w(\alpha_j,\alpha_i)}.
        \end{align*}
Defining $V_i \as V^{(1)}$ and recalling that $V=V_{i-1}$, we 
complete the proof of the inductive claim \refeq{vi1}. Applying the claim
for $i=1$ to $r$, and making appropriate substitutions, we get the
desired relation \refeq{bound0}.
        \beql{bound0}
        \det(V_0 ) = \det(V_r) \prod_{i=1}^r \prod_{\alpha_j \in \In(\alpha_i)}(\alpha_i-\alpha_j)^{w(\alpha_j,\alpha_i)},      
        \eeql
where $V_0 = V(\bfalpha;\bfmu)$ (see \refeq{V0})
The absolute value of the product on the right-hand side is the value that we need to lower bound; 
we know the determinant on the left-hand side from \refPro{det}, so all that remains is to derive
an upper bound on $|\det(V_r)|$. We will use Hadamard's inequality for this purpose, which requires
us to derive an upper bound on the two-norms of the columns of the matrix $V_r$.
Let \dt{$V_r(\alpha_i;j)$ denote the $j$th column of the block of columns $V_r(\alpha_i)$}
corresponding to $B(\alpha_i)$ in $V_0$; note that $V_r(\alpha_i)$ may be the same as
$B(\alpha_i)$ (this happens, for instance, when there are no edges incident on $\alpha_i$ in $\calG$).
In what follows, we derive an upper bound on $\|V_r(\alpha_i;j)\|_2$. 

Recall the definition of $N_j$ from \refeq{Nj}, and
that $\mu_i$ is the size of the block $B(\alpha_i)$.
For convenience again, let the sets $S_j, S_{j+1} \dd S_{\mu_i} \ib \In(\alpha_i)$ be indexed
such that $S_j = \set{1 \dd |S_j|}$, the next $|S_{j+1}|$ numbers are in $S_{j+1}$
and so on until $S_{\mu_i}$ is the last $|S_{\mu_i}|$ numbers smaller than $N_j$;
thus these sets form a partition of the set $\set{1 \dd N_j}$. Now 
the $m$th entry in the column $V_r(\alpha_i; j)$ is \refeq{fmi}.
From \refeq{closedpf}, we have the following bound on the absolute value of \refeq{fmi}
after substituting $n \as N_j+1$, $y_0 = \alpha_0 \as \alpha_i $, 
$y_\ell \as \alpha_\ell$, $\ell =1 \dd N_j$, and 
the indices $i_{\ell}$'s are defined as in \refeq{i0} and \refeq{il}:
        $$\mathlarger{\sum}_{\substack{(t_0, t_1 \dd t_{N_j})\in \ZZ_{\ge 0}^{N_j+1}\\
                                        t_0+t_1+\cdots + t_{N_j} = m-1-N_j}} \prod_{\ell=0}^{N_j} {t_\ell \choose i_\ell}\abs{\alpha_\ell}^{t_\ell-i_\ell}.$$
Since $\alpha_1 \dd \alpha_{N_j}$ have edges directed to $\alpha_i$, their absolute values
are smaller than $|\alpha_i|$. Therefore, the quantity above is upper bounded by
        $$\mathlarger{\sum}_{\substack{(t_0, t_1 \dd t_{N_j})\in \ZZ_{\ge 0}^{N_j+1}\\
                                        t_0+t_1+\cdots + t_{N_j} = m-1-N_j}} \prod_{\ell=0}^{N_j} {t_\ell \choose i_\ell}\abs{\alpha_i}^{m-1-N_j-i_\ell},$$
which is equal to
        \beql{bound1}
        \abs{\alpha_i}^{m-1-N_j-\sum_{\ell=0}^{N_j}i_\ell}\mathlarger{\sum}_{\substack{(t_0, t_1 \dd t_{N_j})\in \ZZ_{\ge 0}^{N_j+1}\\
                                        t_0+t_1+\cdots + t_{N_j} = m-1-N_j}} \prod_{\ell=0}^{N_j} {t_\ell \choose i_\ell}.
        \eeql
Define
        \beql{Mj}
        M_j \as N_j + \sum_{\ell=0}^{N_j} i_\ell = N_j + j-1+\sum_{\ell =1}^{N_j}i_\ell,
        \eeql
where the second equality follows from the fact that $i_0 = j-1$ (see the definition in \refeq{i0}).
The binomial coefficients ${t_\ell \choose i_\ell}$ vanish for $t_{\ell} < i_{\ell}$, so we can assume that
$t_{\ell} \ge i_{\ell}$. If $j_{\ell} \as t_\ell - i_\ell$, then 
        $$\sum_{\ell=0}^{N_j} t_\ell = \sum_{\ell=0}^{N_j} i_\ell + \sum_{\ell=0}^{N_j} j_\ell,$$
and so the constraint $\sum_{\ell=0}^{N_j} t_\ell = m-1-N_j$
is equivalent to 
        $$\sum_{\ell=0}^{N_j} j_\ell = m - 1- N_j - \sum_{\ell=0}^{N_j} = m-1-M_j$$
where the last step follows from the definition of $M_j$ \refeq{Mj}.
Changing the indices from $t_\ell$ to $j_\ell$ in \refeq{bound1}, we get the following
bound the $m$th entry of $V_r(\alpha_i;j)$:
        \beql{bound2}
        \abs{\alpha_i}^{m-1-M_j}\mathlarger{\sum}_{\substack{(j_0, j_1 \dd j_{N_j})\in \ZZ_{\ge 0}^{N_j+1}\\
                                        j_0+j_1+\cdots + j_{N_j} = m-1-M_j}} \prod_{\ell=0}^{N_j} {i_\ell+j_\ell \choose i_\ell}.
        \eeql
We next derive a closed form for the summation term above.

Consider the generating function
        $$\sum_{t_\ell \ge i_\ell} {t_\ell \choose i _\ell}x^{t_\ell-i_\ell} 
                = \sum_{j_\ell \ge 0} {i_\ell+j_\ell \choose j _\ell}x^{j_\ell} = (1-x)^{-(i_\ell+1)}$$
for a given $\ell$. Taking the product of these for different choices of $\ell$, it follows that the summation term in the right-hand side
of \refeq{bound2} is the coefficient of $x^{m-1-M_j}$ in the generating function
        $$(1-x)^{-\sum_{\ell=0}^{N_j}(i_\ell+1)} = (1-x)^{-(M_j+1)}$$
which is 
        $${m-1 \choose M_j}.$$
This implies that \refeq{bound2} is equal to $|\alpha_i|^{m-1-M_j}{m-1 \choose M_j}$.

From the argument in the preceding paragraph,
it follows that in the matrix $V_r$ the two-norm of the $j$th column, in the block 
of columns corresponding to $B(\alpha_i)$, is
        $$\|V_r(\alpha_i;j)\|_2 \le \paren{\sum_{m=1}^{n} |\alpha_i|^{2(m-1-M_j)}{m-1 \choose M_j}^2}^{1/2}.$$
Since for $m-1 \le M_j$ the binomial term vanishes, we can start the summation from 
$M_j$ onwards to obtain the following equivalent form
        $$\|V_r(\alpha_i;j)\|_2 \le \paren{\sum_{m=M_j}^{n-1} |\alpha_i|^{2(m-M_j)}{m \choose M_j}^2}^{1/2}.$$
Substituting $|\alpha_i|$ by 
        \beql{max1}
        \max_1 |\alpha_i| \as \max \set{1, |\alpha_i|}
        \eeql
and pulling out its largest power
from the summation we have the following upper bound on the two-norm
        $$\|V_r(\alpha_i;j)\|_2 \le \max_1|\alpha_i|^{(n-1-M_j)}\paren{\sum_{m=M_j}^{n-1} {m \choose M_j}^2}^{1/2}.$$
Using the upper bound from  \cite[Lemma 7]{perruci} on the summation term above, 
we get the following inequality
        $$\|V_r(\alpha_i;j)\|_2 \le \max_1|\alpha_i|^{(n-1-M_j)} \paren{\frac{n}{\sqrt{3}}}^{M_j} \sqrt{n}.$$
Taking the product of these quantities for $j=1 \dd \mu_i$, we get the following upper bound on the
product of the two-norms of the columns in the block $V_r(\alpha_i)$ in $V_r$:
        \beql{bound3}
        \prod_{j=1}^{\mu_i}\|V_r(\alpha_i;j)\|_2 \le \max_1|\alpha_i|^{\sum_{j=1}^{\mu_i}(n-1-M_j)} \paren{\frac{n}{\sqrt{3}}}^{\sum_{j=1}^{\mu_i} M_j} {n}^{\mu_i/2}.
        \eeql
Let us understand the term $\sum_{j=1}^{\mu_i}M_j$.

\bleml{mj}
For a vertex $\alpha_i$ in the directed acyclic graph $\calG$, define
        \beql{wi}
        w_i \as \sum_{\alpha_\ell \in \In(\alpha_i)}w(\alpha_\ell,\alpha_i),
        \eeql
that is, the sum of the weights of all edges incident on $\alpha_i$.
 Then
        $$\sum_{j=1}^{\mu_i}M_j = {\mu_i \choose 2} + w_i.$$
\eleml
\bpf
Recall the definition of the sets $S_j$, from \refeq{sj}, and the definition of $M_j$, from \refeq{Mj}. 
Given a $j$, and  $\ell \in S_j$, $i_\ell = r_\ell-1$ from \refeq{i0}; for $\ell \in S_\kappa$, where $\kappa = j+1 \dd \mu_i$,
$i_\ell=\mu_\ell-1$. Therefore, we can rewrite
\refeq{Mj} as
\begin{align*}
        M_j &= N_j+j-1 + \sum_{\ell \in S_j}(r_{\ell}-1) + \sum_{\ell \in \su_{\kappa > j}S_\kappa} (\mu_\ell-1)\\
               &= j-1 + \sum_{\ell \in S_j}r_{\ell} + \sum_{\ell \in \su_{\kappa > j}S_\kappa} \mu_\ell.
\end{align*}
The sum $\sum_j \sum_{\ell \in S_j}r_\ell$ is the sum of the residue terms over all indices in $\su_{j=1}^{\mu_i}S_j$. 
Now consider the sum
        $$\sum_{j=1}^{\mu_i} \sum_{\ell \in \su_{\kappa > j}S_\kappa} \mu_\ell.$$
For two indices $j < \kappa$, the summation over $j$ contributes an $\mu_\ell$ for every $\ell \in S_\kappa$.
Therefore,
        $$\sum_{j=1}^{\mu_i}\paren{\sum_{\ell \in S_j}r_{\ell} + \sum_{\ell \in \su_{\kappa > j}S_\kappa} \mu_\ell} = w_i$$

\epf

Substituting the result in the lemma above into \refeq{bound3}, we get the following upper bound on the
two-norms of the columns  in $V_r(\alpha_i)$
        $$
          \prod_{j=1}^{\mu_i}\|V_r(\alpha_i;j)\|_2 \le
          \max_1|\alpha_i|^{(n-1)\mu_i - {\mu_i\choose 2}-w_i} \paren{\frac{n}{\sqrt{3}}}^{{\mu_i \choose 2} + w_i} {n}^{\mu_i/2}.
        $$
Taking the product of this bound for $i=1 \dd r$, along with Hadamard's inequality, 
gives us the following upper bound
        \beql{bound5}
        |\det(V_r)|\le\prod_{i=1}^r\paren{\max_1|\alpha_i|^{(n-1)\mu_i - {\mu_i\choose 2}-w_i} \paren{\frac{n}{\sqrt{3}}}^{{\mu_i \choose 2} + w_i}} {n}^{n/2}.
        \eeql
where we use the fact that $n=\sum_{i=1}^r \mu_i$.
The term
        $$
        (n-1)\mu_i - {\mu_i\choose 2}-w_i 
        = \sum_{j=1; j \neq i}^r \mu_i \mu_j - w_i + {\mu_i \choose 2}
        < \sum_{\substack{j=1\\ j \neq i}}^r \mu_i \mu_j - w_i + \mu_i^2.
        $$
If $\bfmu$ be the
column vector of all $\mu_i$'s, 
and $A_w$ be the adjacency matrix with the $(i,j)$th entry as the weight 
$w(\alpha_i, \alpha_j)$ of the corresponding edge $(\alpha_i,\alpha_j)$, then
the last term in the inequality above is the one-norm of the $i$th row of the matrix $\bfmu \bfmu^t - A_w$.
Since the $\infty$-norm of the matrix $\| \bfmu \bfmu^t - A_w\|_\infty$ is the maximum over all the row-sums,
we have
        $$(n-1)\mu_i - {\mu_i\choose 2}-w_i  \le \| \bfmu \bfmu^t - A_w\|_\infty.$$
As for the term
        $$\sum_{i=1}^r\paren{ {\mu_i \choose 2} + w_i} = \sum_{i=1}^r {\mu_i \choose 2} + w(E),$$
where $w(E)$ is defined in \refeq{we}.
Substituting these bounds in \refeq{bound5}, we obtain the following upper bound
        \beql{bound6}
        |\det(V_r)|\le M(\bfalpha)^{\| \bfmu \bfmu^t - A_w\|_\infty} \paren{\frac{n}{\sqrt{3}}}^{\sum_i {\mu_i \choose 2} + w(E)} {n}^{n/2}.
        \eeql
Substituting this upper bound in \refeq{bound0} and moving it to the denominator in the left-hand side
completes the proof of  \refThm{main}.

\subsection{Choosing the best matrix}
\label{sec:best-matrix}

\refThm{main} leaves open the choice of the potentials $\mu_i \in \NN$, $i=1 \dd r$.
Our aim here is to find the best possible choice of $\mu_i$'s satisfying the edge constraints 
$w(\alpha_i,\alpha_j) \le \mu_i\mu_j$ and at the same time 
minimizing $\|\bfmu \bfmu^t - A_w\|_\infty$. 
For example, if all the weights are one then it is clear that
$\mu_i=1$, for $i=1 \dd r$, is the best possible assignment. In which case,
        $$V(\bfalpha; \bfmu) = V(\bfalpha),\;
        \|\bfmu \bfmu^t - A_w\|_\infty\le (r-1),\; n=r,\; w(E) = |E|$$
 and so \refThm{main} matches the bound given in \refPro{perruci}. 

Consider the relaxed version of the problem where $\mu_i$'s are positive reals; it is clear that rounding them
up to the nearest integer would give a valid solution (though not an optimum solution)
to the problem over the positive integers.
Then the optimization problem is
to minimize  $\|\bfmu \bfmu^t - A_w\|_\infty$ such that
        $$\bfmu \bfmu^t \ge A_w$$
where `$\ge$' here means entry wise; note that the non-edge constraints are trivially satisfied
since no $\mu_i$ is ever assigned to zero. Since $A_w$ is non-negative, we know
from the Perron-Frobenius theory \cite{hj} that the 
spectrum of $A_w$ is an eigenvalue $\rho(A_w)$ of $A_w$. 
Moreover, as $A_w$ is symmetric it can be orthogonally diagonalized,
i.e., $A_w = Q \Lambda Q^t$, where $Q$ is the $r\times r$ orthogonal matrix whose columns
$\bfq_k$, $k=1 \dd r$, are the eigenvectors of $A_w$ and $\Lambda$ is a diagonal matrix
that has the corresponding eigenvalues of $A_w$. Another way to express the  relation is that
$A_w$ is the sum of some rank one matrices obtained by its eigenvectors, i.e., 
        $$A_w = \sum_{k=1}^r \lambda_k \bfq_k \bfq_k^t.$$
We can also assume that the $\|\bfq_k\|_2=1$ for $k=1 \dd r$. Combined with
 the equation above it follows that the $(i,j)$-th entry of $A_w$ 
        $$w(\alpha_i,\alpha_j)=\sum_{k=1}^r \lambda_k \bfq_{k,i}\bfq_{k,j}.$$  
Since by assumption $\|\bfq_k\|_2=1$, taking absolute values we get
        $$w(\alpha_i,\alpha_j) \le \sum_{k=1}^r |\lambda_k|  = \|A_w\|_\star,$$
where $\|A_w\|_\star$ is the \dt{nuclear norm} of $A_w$.
Therefore, we can take $\bfmu$ in \refThm{main} as the vector
        \beql{mu}
        \bfmu \as \ceil{\sqrt{\|A_w\|_\star}} \overbrace{(1, 1 \dd 1)}^{r},
        \eeql
which implies that 
        $$n = r\ceil{\sqrt{\|A_w\|_\star}}$$
in the theorem. The error in the approximation can be shown to be bounded
by
        \begin{align*}
          \|\bfmu\bfmu^t - A_w\|_\infty 
          &\le 2r \|A_w\|_\star,
        \end{align*}
and
        $$\sum_i {\mu_i \choose 2}  
        \le \frac{3r \|A_w\|_\star}{2},$$
where in the last inequality we use the observation that as $A_w$ has entries in $\ZZ_{\ge 0}$,
its spectrum is greater than one, and hence $\|A_w\|_\star \ge 1$.
By making these substitutions in \refThm{main}, we obtain the result, namely \refeq{result1}, 
mentioned in \refSec{intro}.

\section{Conclusion and Future Work}
\label{sec:conclusion}
Our derivation using the confluent Vandermonde matrix to get the desired weights in
the exponents has the advantage of optimizing over the various choices of the matrix.
We have given a first attempt at exploiting this choice. Whereas rank-one approximations
to matrices are well studied \cite{friedland:tensor-rank-one:13}, the challenge in our context 
is to derive a symmetric rank-one matrix that also {\em dominates} $A_w$. 

One would also like
to derive a lower bound on the absolute value of  $\det(V(\bfalpha; \bfmu))$ in terms of
the polynomial $f$, to get a more direct comparison with the earlier results. Perhaps
an algorithm to compute the determinant from the coefficients would also be interesting;
a related recent result is an algorithm to compute the $D^+(f)$-root function defined as
$\prod_{1 \le i < j \le r}(\alpha_i- \alpha_j)^{m_i + m_j}$, i.e., $G$ is the complete graph on
the roots and the weight of an edge is the sum of the multiplicity of its vertices \cite{yang-yap}.
Similar to \cite{emt}, one would like to derive weighted version of the results
for the more general setting of polynomial systems.



\myTail